\documentclass[preprint]{emulateapj} 

\usepackage{natbib}
\usepackage{hyperref}
\hypersetup{colorlinks,citecolor=Blue,linkcolor=Red,urlcolor=Blue}
\usepackage{amsmath}
\usepackage{empheq}
\usepackage[usenames,dvipsnames]{color}
\allowdisplaybreaks 

\begin{document}
 
\title{Orbital clustering in the distant solar system}
\author{Michael E. Brown \& Konstantin Batygin} 

\affil{Division of Geological and Planetary Sciences, California Institute of Technology, Pasadena, CA 91125} 
\email{mbrown@caltech.edu, kbatygin@gps.caltech.edu}

\begin{abstract}
The most distant Kuiper belt objects appear to be clustered in
longitude of perihelion and in orbital pole position. To date, the only
two suggestions for the
cause of these apparent clusterings have been either the effects of observational bias 
or the existence of the distant giant planet in an eccentric
inclined orbit known as Planet Nine. To determine if observational
bias can be the cause of these apparent clusterings, we develop a rigorous
method of quantifying the observational biases in the observations of
longitude of perihelion and orbital pole position. From this now more
complete understanding of the biases we calculate that the probability that
these distant Kuiper belt objects would be clustered as strongly
as observed in both longitude of perihelion and in orbital pole position
is only 0.2\%. While explanations other than Planet Nine may 
someday be found, the statistical significance of this clustering is now 
difficult to discount.
\end{abstract}

\section{Introduction}
The apparent
physical alignment of the perihelion position and orbital poles of many of the
most distant Kuiper belt objects
(KBOs) has been taken to be evidence of a massive eccentric inclined 
planet well beyond Neptune \citep{2016AJ....151...22B}. 
To date no successful 
alternative dynamical explanation for such clustering has been
suggested. Instead, discussion has focused on the possibility that
the apparent physical alignments might be due to either random chance
or highly affected by observational bias or both. Observational bias
for distant objects is of critical concern. Objects on these extremely
eccentric orbits are
often found close to perihelion, so the sky distribution of 
observational surveys has a strong effect on the distribution of
observed values of longitude of perihelion. Observations near the 
ecliptic preferentially find objects with longitude of ascending
node close to the longitude of the observation, providing a strong
potential bias to the pole position, too.

In \citet{2016AJ....151...22B} we made a simple estimate of the 
possibility of observational 
bias by suggesting that the biases in the discoveries of 
objects with semimajor axes beyond 230~AU should
not differ significantly from those of discoveries from 50 to 230 AU. We found that under this assumption the probability that the clustering
of the then 6 known objects beyond 230~AU was highly significant. 
Given this simplistic initial assumption about observational biases
and the importance of understanding this potential clustering,
\citet{2017AJ....154...65B} developed a more rigorous method of estimating the longitude of perihelion
bias by using the collection of all reported 
KBO discoveries to back out
the probabilities that distant eccentric objects could have been detected
in the collection of all surveys to date. No reliable method was available for evaluating the biases in pole position, however, rendering this work 
incomplete.

Conversely,  calculation of the full 
set of observational biases
for the well calibrated OSSOS survey found that their discovery of
4 objects with semimajor axis beyond 230 AU was consistent with a 
uniform distribution of orbital angles 
\citep{2017AJ....154...50S}, in apparent conflict with the
clustering observed elsewhere. Unfortunately, although OSSOS
collected by far the largest number of well characterized discoveries of
KBOs, the survey was limited to two regions of ecliptic longitude,
and these two longitudes are
nearly orthogonal to the apparent longitude of perihelion clustering. 
As a result, the strength of the constraint on clustering from OSSOS is not yet clear, 
and will be analyzed below.

The lack of an explicit calculation of the bias in pole position 
in  \citet{2017AJ....154...65B} remains a key impediment to a full treatment
of the effect of observational biases on the observed clustering of
distant objects.
Here we extend the technique developed in that paper (which is in
itself an extension of the technique developed in \citet{2001ApJ...554L..95T})
to develop
a method  to fully include both
longitude of perihelion and orbital pole position in our bias calculations. 
We use these full bias calculations to determine the probability that a randomly distributed
set of distant objects would be simultaneously clustered in longitude
of perihelion and in pole position as strongly as the observations suggest. 

\begin{figure*}
 \plotone{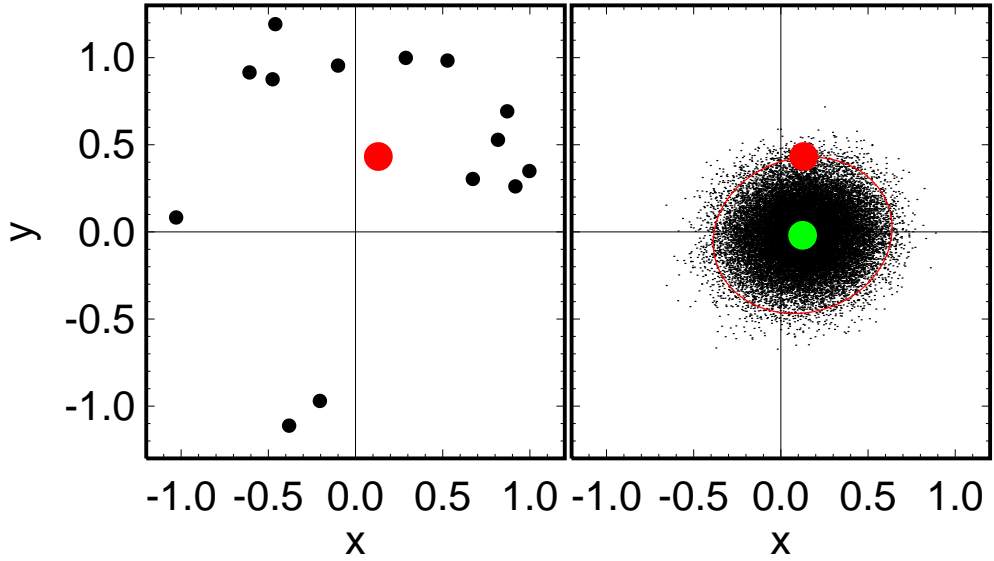}
\caption{(a) The canonical $(x,y)$ coordinates for the
14 known KBOs with semimajor axis greater than 230 AU. These coordinates
point in the direction of the longitude of perihelion of each object but are modulated 
by a function of eccentricity. The average position of $(0.13,0.43)$ is marked
in red. The distance away from the origin of this average is a measure of the strength of the clustering. 
(b) Average $(x,y)$ positions of each of 100,000 iterations in which
we randomly select  longitudes of perihelion from the bias 
distribution functions for each of the 14 objects 
(under the assumption that the longitudes 
are distributed uniformly) and then compute the average $(x,y)$ 
position of the 14 random objects of each iteration. 
The average $(x,y)$ position of the 14 real distant objects
(red point) is more strongly clustered than 96\% of the random iterations
where the longitudes are 
assumed to be uniform (enclosed within the red ellipse). 
Overall, a population of objects which
was uniform in longitude of perihelion would be biased
on average to have a longitude of perihelion clustering 
towards -7 degrees, shown as a green
point.}
\end{figure*}

\begin{figure*}
\plotone{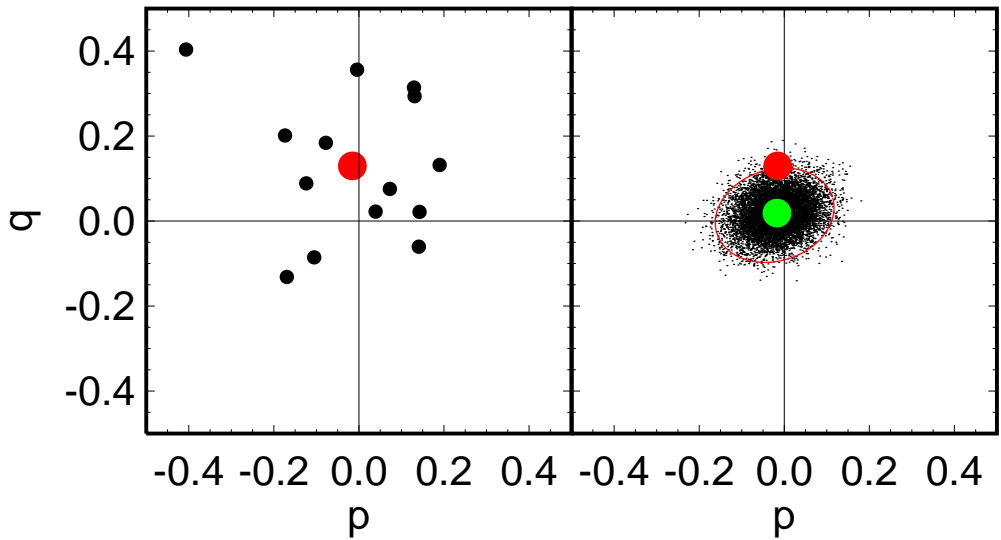}
\caption{(a) The canonical $(p,q)$ coordinates for the
14 known KBOs with semimajor axis greater than 230 AU. These coordinates
point in the direction of the longitude of ascending node with a 
magnitude of approximately $\sin i$, but modulated 
by a function of eccentricity. The average position of $(-0.02,0.13)$ 
is marked
in red. (b) Average $(p,q)$ positions of 100,000 iterations in which
we randomly select pole positions for the 14 distant KBOs
using the bias distribution functions and
assuming a uniform distribution of node and
a scattered disk-like distribution in inclination and calculate the average $(p,q)$ position for each iteration. 
The average $(p,q)$ position of the 14 real distant objects (red point) is more
strongly clustered than 96.5\% of the random iterations (enclosed in the red ellipse). Overall the observations are biased
to have a pole position bias along a line of nodes from 17 to 197 degrees as can be seen from the elongation of the cloud of points along that line.}
\end{figure*}

\section{Longitude of perihelion bias}
Before developing our general method, we first revisit 
and update the longitude of perihelion bias. As of 1 July 2018,
there are 14 known objects with perihelion beyond Neptune and
semimajor axis beyond 230 AU. We calculate the longitude of
perihelion bias for each object independently using the method
of \citet{2016AJ....151...22B}, in which  we use the entire MPC data base 
of reported Kuiper belt object discoveries (again, as of 1 July 2018) 
to determine probabilities of discovering distant KBOs as a function
of longitude of perihelion.
Previously,
we calculated the significance of the clustering by determining the
Rayleigh $Z$ statistic, taking each longitude of perihelion, 
projecting a unit vector in that direction,
taking the two-dimensional average of all of the vectors, and assessing the
significance of the length of the mean vector. Here we slightly modify
this procedure. 

Instead of projecting a unit vector into the geometrical direction 
of the longitude, we instead employ canonically conjugated variables to
more properly represent the orbital parameters. 
Starting with the reduced Poincar\'e action-angle coordinates 
\citep[e.g.][]{Morby_book}
\begin{align}
&\Gamma = 1-\sqrt{1-e^2} &\gamma=-\varpi \nonumber \\
&Z=\sqrt{1-e^2}\, \big(1-\cos (i) \big) &z=-\Omega.
\end{align}
(here 
we have scaled $\Gamma$ and $Z$ by $\Lambda=\sqrt{G\,M_{\odot}\,a}$, as
this factor does not affect our analysis),
we define our clustering in terms of 
canonical Cartesian analogs of Poincar\'e variables, defined as:
\begin{align}
&x = \sqrt{2\,\Gamma}\cos(\varpi) &y = \sqrt{2\,\Gamma}\sin(\varpi) \\
&p=\sqrt{2\,Z}\cos(\Omega) &q=\sqrt{2\,Z}\sin(\Omega).
\end{align}

In practice, the $(x,y)$ coordinate of an object is simply a vector in
the direction of the longitude of perihelion, properly scaled by a function
of eccentricity, and the $(p,q)$ coordinate is a projection of 
the orbital pole position of the object (for small values of $i$)
also properly scaled by a function
of eccentricity. We will thus use the clustering of these vectors
as the appropriate measures for the clustering of the longitude of perihelion and the pole position.
As all of the eccentricities of the distant objects are between 0.69 and 0.98,
using the $(x,y,p,q)$ coordinate system rather than the simple angle of
the longitude of perihelion and the projected pole position
makes little difference to the final result,
but an advantage of adopting these coordinates is that 
by virtue of the Poisson bracket criterion, $(x,y,p,q)$ provides 
an orthogonal basis set for representing orbital parameters.
Below, we will combine the longitude of perihelion and pole position
clustering into a single value in $(x,y,p,q)$ space, necessitating this
orthogonal basis.

With the appropriate variables defined, we
return to computing the bias in the longitude of perihelion. 
Figure 1(a) shows the position of each of the 14 known objects in $(x,y)$
space. The average of these position is (0.13,0.43), corresponding
to an average clustering in longitude of perihelion in the direction of $\varpi=73$ degrees.
The distance from the origin of this average position shows the strength of the clustering,
in analogy to the Rayleigh Z test, though because these are not unit vectors we cannot use the Z test and 
must develop a modified
measure of significance. In order to calculate the significance of 
this observed clustering, then,
we perform 100,000 iterations in which we select a random longitude of
perihelion for each of the 14 objects, sampling 
from the calculated bias distribution
for each object. For each iteration we then calculate the average 
$(x,y)$ coordinate of the 14 randomly selected objects (Fig 1b). This distribution of average positions from the 100,000 iterations shows the probability
that observations of a population that was uniformly distributed in longitude of perihelion
would have a given average clustering.

From the distribution of these
random iterations it is apparent that there is an average bias of $\varpi$
towards $\sim -7$~degrees, nearly perpendicular to the actual clustering 
detected. Because the bias is not symmetric with $\varpi$, we calculate the probability that the clustering of 
the real objects would be as strong as seen and in the direction observed
by finding the minimum-area two-dimensional ellipse which encloses 
the maximum number of random iteration averages and also the real data point.
The red ellipse in Fig 1b encompasses 96\% of the random iterations.
We find that only 4\% of random iterations are clustered as 
significantly in longitude of perihelion as the real data. 

\begin{figure}
\plotone{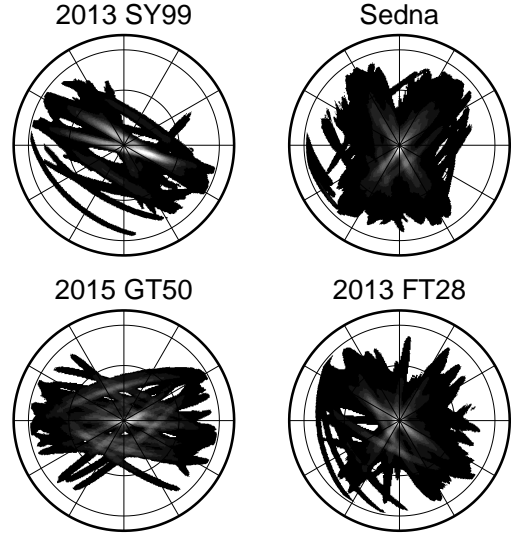}
\caption{The probability of detection of a
sibling KBO to a distant KBO, as a function of the sibling orbital pole position, shown as a polar
projection. Radial grid lines
are shown ever 30 degrees in ascending node, with $\Omega=0$ at the 
right, and grid circles are 
shown every 30 degrees of inclination, with zero inclination at the center. }
\end{figure}

\begin{figure*}
\plotone{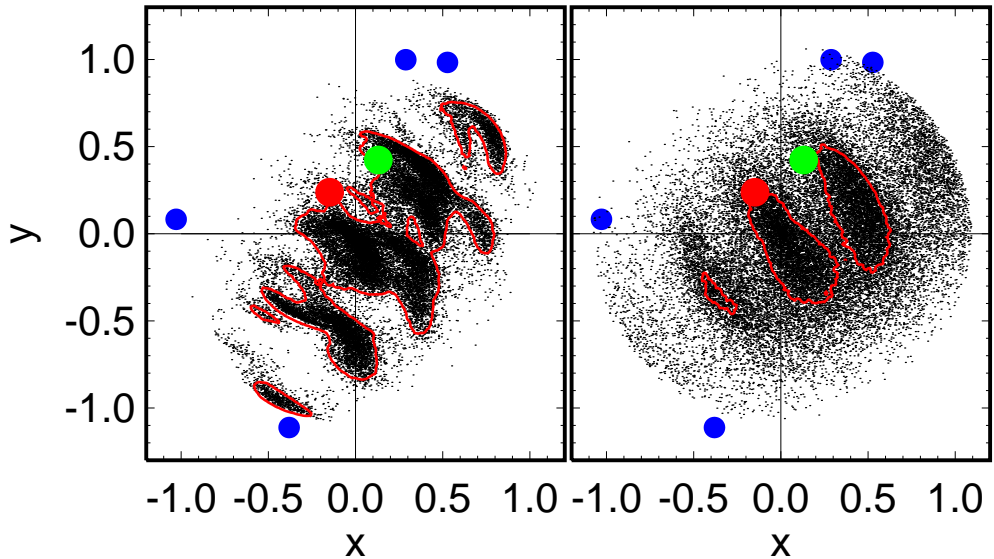}
\caption{The canonical $(x,y)$ coordinates for the 4 distant OSSOS discoveries (blue dots). The mean of
the OSSOS discoveries is shown as a red dot, while the mean of the full set of 14 objects is shown as the 
green dot. (a) 100,000 iterations of randomly sampling our derived OSSOS biases and calculating a mean $(x,y)$
value. The mean values are patchy because there are only 4 objects and the biases are strong in two opposite 
directions.The red contours enclose 85\% of the points and intersect the OSSOS mean. Only 15\% of the 
random iterations are more clustered than the real OSSOS objects, an approximately 1$\sigma$ effect.
(b) 100,000 iterations of randomly sampling using the published average OSSOS biases, rather than constructing
and object-by-object bias. The large spread in longitude of
perihelion positions (compared to Fig 1(b), for example) shows the relative insensitivity of OSSOS to longitude of
perihelion clustering compared to using the full dataset. A clustering as strong as detected in the full
data set (green dot) could only be detected at the 65\% confidence level using the OSSOS survey. Because of the
limited survey region and small number of detected distant objects, OSSOS observations are equally consistent with
being drawn from a uniform distribution of longitudes of perihelion or with being clustered in longitude of
perihelion as strongly as seen in the ensemble data.
}
\end{figure*}

\section{Orbital pole bias}
The distant KBOs also appear to be clustered in orbital pole position. 
In Figure 2a we show the positions of the 14 known distant objects in
$(p,q)$ space, which essentially shows the orbital pole positions of
the objects. The average position of $(-0.02,0.13)$ is marked. 

As with the longitude of perihelion debiasing,
a full debiasing of the observations to account for pole position
would require detailed knowledge 
of all surveys which have detected KBOs. With rare exception, such 
information is not available. Instead we simply have lists of reported
objects which include the derived orbital parameters, brightness,
and detection location of the object. We extend the technique
developed in \citet{2017AJ....154...65B} to use this information to compute an object-by-object
debiasing of the pole positions and inclinations of the distant objects.

We use the fact that a survey which discovered a KBO with a particular
brightness at a particular location would have been equally sensitive to
any KBO at least as bright as the detected object if it had been
at that position at that time (we discuss important
caveats to this assumption below). 
Our procedure is thus as follows. For each individual distant 
KBO, we go through every known discovered KBO and ask which
siblings of the distant  KBO could have been discovered at the
position of the discovered KBO. We define the siblings to be all potential
KBOs with the same absolute magnitude,
$a$, $e$, and $\varpi$ as the distant  object, but having the
same sky position as the KBO discovery. We allow the siblings to be
uniformly distributed in direction of orbital motion around 
the discovery position.
The values of $\Omega$, $i$, and mean anomaly, $M$ for the sibling
are uniquely defined by the sky position of the discovered KBO and its
direction of orbital motion. For 
the derived value of $M$ we calculate the heliocentric distance
and thus the relative magnitude that
the sibling KBO would have at that location. If the sibling KBO is
bright enough that it would have been discovered at that location
(that is, it is as bright or brighter than the true discovered KBO),
we add one to the cell position of an equal area grid on the sky
corresponding to the pole of the sibling KBO (implemented using the 
IDL HEALpix library\footnote{https://healpix.jpl.nasa.gov/html/idl.htm}).
Repeating this process
for all detected KBOs builds a map that shows the relative probability
of detection of the distant KBO under the assumption that pole positions
are uniform across the sky. We then repeat the entire process for each
individual distant KBO to build individual probability distributions.

This debiasing method relies on the assumption that KBOs discovered are a fair representation
of the surveys that have been carried out and that the surveys could have detected any sibling KBO brighter than
the real detections. Important caveats and corrections to these assumptions, including
corrections for KBO distance, longitudinal clustering of resonant KBOs, and the latitudinal surface density of
KBOs are discussed in \citet{2017AJ....154...65B}. We test the sensitivity
of the results to each of these assumptions and find that the results
do not depend strongly on any of the precise assumptions.
We add one modification to that analysis.
The large survey of \citet{2016ApJ...825L..13S} is dedicated to finding distant objects. 
As such, only objects with initial discovery distances greater than 50 AU are tracked. 
The actual survey, then, covers a much larger area 
than would be inferred from the reported discoveries. We thus weight
each object discovered by \citet{2016ApJ...825L..13S} by the ratio of the 
expected number of objects that they would have discovered had they tracked 
the closer objects to the number of distant objects they actually detected. 
In detail, we determine this ratio by noting that
the number density
of scattered disk objects (which comprise a majority of objects with discovery distance $r>50$ AU)
as a function of latitude is approximately proportional to the number density
of hot classical KBOs as a function of latitude \citep{2001AJ....121.2804B}, so the
appropriate weighting is the total number of known objects 
(with $a>40$ and $r>30$ AU, as discussed in \citet{2017AJ....154...65B}) discovered
at absolute heliocentric latitudes greater than 5 degrees (to exclude the cold classical
objects) divided by the total number of these objects discovered at these latitudes
but with $r>50$ AU, in both cases excluding objects discovered by the Dark Energy Camera
(DEC) which was used for the survey (note that {\it all} DEC discoveries are excluded 
because it is not possible to always determine which specific survey made a 
specific discovery). We find 362 total objects with $r>30$ AU and 35 with $r>50$ AU, for a
ratio of $10\pm2$. The results are again insensitive to the precise value used.

The probability distribution functions for the pole positions of four representative
distant KBOs are shown
in Figure 3. 
The pole positions of discoveries of distant KBOs are heavily
biased. Deep limited area surveys like OSSOS and the Dark Energy Survey
contribute specific high probability bands in the maps that are easily
identified. Surveys that cover the sky more uniformly have their probabilities
more distributed and, while they contribute heavily to the overall probability 
distribution, they do not stand out as clearly.

While the biases are severe, no clear bias towards a pole offset
in the observed direction is obvious.
We assess the likelihood
that observational biases lead to the clustering in pole position
by performing 100,000 iterations of a test where we choose a random object
from the inclination distribution-weighted probability distribution of each of the 14 objects and then
calculate the average $(p,q)$ coordinates for each iteration.

As with the longitude of perihelion clustering, the observed clustering in $(p,q)$ space
of the 14 distant object appears stronger than most of the iterations in which we assume
a uniform distribution of longitudes of ascending node. To determine the
significance of this clustering, 
we again fit a minimum area two-dimensional ellipse to maximize the number
of included randomly chosen $(p,q)$ points and to also include the real observed clustering.
The red ellipse in Fig 2b encloses 96.5\% of the random points. 
Only 3.5\% of
the random iterations are as strongly clustered in pole
position as the real observations.
As can be seen (Fig~2b), a bias exists towards
pole positions with longitudes of ascending nodes along a line from $\sim 17$ to $\sim 197$~degrees, the approximate 
longitudes of the OSSOS survey regions. 
\section{Combining longitude of perihelion and pole position bias}
The probability that both longitude of perihelion and pole position would be clustered
is significantly lower than that of either one independently. 
We use our previously developed method to extend our bias analysis to
explicitly calculate the probability of finding an object with both
a given longitude of perihelion and pole position. 
To perform this extension we simply take our pole position
bias method, which required that the object have a longitude
of perihelion equal to that observed, and recalculate the pole position
bias as a function of a (assumed to be uniform) longitude of perihelion,
which we discretize into one degree bins. For each of the 14 observed objects,
with observed values of $a$, $e$, and $H$, we
now explicitly have the probability distribution function that
such an object would have been discovered
with a given longitude
of perihelion and pole position under the assumption that the underlying
distribution is uniform in longitude of perihelion and in pole position.

While a uniform longitude of perihelion is the correct assumption against
which to test, we clearly do not want to assume a uniform distribution
of poles. Instead we assume a uniform distribution in longitude of
ascending node and seek a realistic distribution in inclination.
We examine the inclinations
of distant KBOs using the method developed by \citet{2001AJ....121.2804B}. 
Perhaps the most appropriate inclination distribution to assume
for our objects would be the inclination distribution of the 
entire scattered disk. We find that the 149 multi-opposition
objects with $a>50$ AU and $q>30$ AU are fit by a standard function
of $f(i)\propto \sin(i) \exp(-i^2/2\sigma^2)$ with $\sigma=14.9\pm 0.6$ degrees. It is
possible that the more distant objects have a different inclination
distribution, however. If we fit only the 27 multi-opposition objects
with $a>150$ AU and $q>30$ AU, we find a best-fit of $\sigma=15\pm 2$ degrees, consistent
with the larger sample. Finally we examine just the 14 objects with $a>230$ AU
and $q>30$ AU. Here we find a slightly lower inclination distribution of
$\sigma=11^{+3}_{-2}$ degrees. In the significance analysis below, a narrower inclination
distribution will make a given pole position clustering appear
more significant, as a narrow inclination distribution is less likely to
have an average far from the ecliptic pole. Thus, to be conservative, 
we use $\sigma=16$ degrees - larger than any of the best fit values -- 
as our assumed inclination distribution.

With the full probability distribution calculated, we now perform 1.6 million
iterations where we randomly choose a longitude of perihelion
and a pole position from the probability distribution function
for each of the 14 objects and calculate the average
of those 14 objects in $(x,y,p,q)$ space. We perform a four-dimensional
analog to the ellipse fitting of Figs. 1 \& 2 and find that only 99.8\% of
iterations have an average as extreme at the real measured values
of the distant objects. This number corresponds to a probability 
of finding the combined longitude of perihelion and pole cluster of
only 0.2\%.

\section{Comparison to OSSOS}
The OSSOS observations of 4 distant KBOs have been suggested to be consistent with
a uniform distribution of longitudes of perihelion \citep{2017AJ....154...50S}.
We examine whether the severe observational selection
effects of OSSOS coupled with the small number of distant objects discovered
might prevent them from detecting the longitude
of perihelion clustering
seen here. 

Figure 4 shows the $(x,y)$ positions of the 4 OSSOS objects (blue points).
These objects have an average $(x,y)$ of $(-0.14,0.24)$, corresponding
to an average longitude of perihelion of $140$ degrees 
(red point), similar to
the average $(x,y)$ value from the full dataset of 14 objects (green point). 
To understand the biases for these objects from the OSSOS survey
alone we use the method
developed above.
We create longitude of perihelion bias
distributions for the 4 distant OSSOS discovered objects using
only the KBOs discovered in the OSSOS survey as our observational set. Because
of the limited range of ecliptic longitudes of OSSOS the biases for the individual objects are
significantly more severe than found when using the complete catalog set.
We then perform 100,000
random iterations as above, where we randomly select longitudes of
perihelion for
the 4 OSSOS objects from the bias distribution develeoped from the full
set of OSSOS objects. Figure 4a shows the average $(x,y)$ position
of the 4 objects from each of the 100,000 iterations.
Where similar results for the full dataset showed a smooth distributions
of points (Fig 1b), the distribution here is quite clumpy. This
clumpiness is driven by the strong biases in the potential discovery
longitudes of perihelion in the OSSOS survey and the small number of
objects being averaged. In spite of this clumpiness, it can be seen that
a strong bias exists for the 
average longitude of perihelion to be somewhere along an axis from
$\sim$45 degrees to $\sim$225 degrees.
Interestingly, the mean $(x,y)$ position of the 4 OSSOS discoveries
is displaced in the {\it orthogonal} direction to the strong biases. 
A two-dimensional ellipse is a poor representation of the 
spread of the data in this case,
so we instead draw contours enclosing the regions of highest density.
We find that only 15\% of the random iterations are more
strongly clustered than the four real objects, an approximately 1$\sigma$ effect. These four objects cluster
approximately in the same direction as the clustering seen in the 
full dataset. 

\citet{2017AJ....154...50S} calculate biases differently. Rather than
evaluate the biases on an
object-by-object basis, considering the specific orbital elements of each
object, they instead assume a specific distribution for the orbital 
elements of distant 
objects and generate biases for this assumed distribution. We evaluate 
how this alternate method affects the longitude of perihelion bias. We use
the reported OSSOS 
derived average bias function of \citet{2017AJ....154...50S} and repeat our
analysis with a separate set of average $(x,y)$ positions found in
100,000 iterations (Fig 4b). This average set of $(x,y)$ positions 
is smoother, as expected from using a single average bias, though with
approximately the same overall distribution as above. Here we find that 
the OSSOS observed clustering
is no longer significant, as the random spread in longitude of 
perihelion has increased to be larger than the 
measured clustering. More importantly we find that even if 
a clustering as strong as that 
observed in the full data set were present (green point in Fig. 4), 
it could only be detected at the 65\% confidence level. That is,
the uncertainties in the measurement of 
clustering from the OSSOS data are so large that OSSOS would not be
capable of confidently detecting the clustering seen in the larger data set
even if it were real and present in the OSSOS data. 
Because of the limited survey region and small
number of detected objects, OSSOS observations are equally consistent 
with being drawn from a uniform 
distribution of longitudes of perihelion and with being clustered 
in longitude of perihelion as strongly
as seen in the ensemble data. No conclusions on clustering of longitude 
of perihelion observed in the complete dataset can be drawn from 
the OSSOS data.

\section{Conclusions}
Fully taking into account the biases in longitude of perihelion, inclination,
and longitude of ascending node, the probability that the 14 known KBOs with
semimajor axes beyond 230 AU would be clustered as strongly as
currently observed due only to a combination of observational bias and
random chance is only 0.2\%.  
While it remains true that the OSSOS survey is consistent with
a uniform distribution of these parameters, we have shown that the
severe longitudinal bias of the OSSOS survey renders it insensitive to
the clustering observed in the more evenly distributed surveys.
While the existence of Planet Nine remains the only current hypothesis
for the explanation of this clustering \citep{2016AJ....151...22B}, 
we have shown 
that the joint clustering in longitude of perihelion and pole position
of the distant KBOs is nearly indisputable, 
regardless of the existence of Planet Nine.
If Planet Nine is not responsible for this clustering new dynamical
processes need to be found in the outer solar system. 

We would like to thank the referee for an insightful question which
lead to a significant improvement in this analysis. Discussions with
David Gerdes, Matt Holman, Chad Trujillo, and Elizabeth Bailey helped
to shape these arguments.

\end{document}